\documentclass[lettersize,journal]{IEEEtran}
\usepackage{amsmath,amsfonts}
\usepackage{algorithmic}
\usepackage{algorithm}
\usepackage{array}
\usepackage{hyperref}
\usepackage[caption=false,font=normalsize,labelfont=sf,textfont=sf]{subfig}
\usepackage{textcomp}
\usepackage{stfloats}
\usepackage{url}
\usepackage{orcidlink}
\usepackage{verbatim}
\usepackage{graphicx}
\usepackage{cite}
\usepackage{booktabs}
\usepackage{tabularx}
\begin{document}

\title{NeuroCodec: EEG-Guided Latent-Domain Generative Model for High-Fidelity Target Speaker Extraction}


\author{Julian~Silva\,\orcidlink{0009-0009-3542-5723},
        Jaliya~Kodithuwakku\,\orcidlink{0009-0006-9953-5217},
        Navindu~Sirimanna\,\orcidlink{0009-0002-9404-2289},
        Shaveen~Herath\,\orcidlink{0009-0008-2299-601X},
        Wageesha~N.~Manamperi\,\orcidlink{0000-0002-5864-4154},~\IEEEmembership{Member,~IEEE},
        and~Joshua~P.~Kulasingham\,\orcidlink{0000-0003-3599-9160},~\IEEEmembership{Member,~IEEE}% <-this % stops a space
\thanks{J. Silva, J. Kodithuwakku, N. Sirimanna, S. Herath, W. N. Manamperi, and J. P. Kulasingham are with the Department of Electronic and Telecommunication Engineering, University of Moratuwa, Katubedda 10400, Sri Lanka (e-mail: silvaljjp.21@uom.lk; kodithuwakkujn.21@uom.lk; sirimannantw.21@uom.lk; herathbhmksb.21@uom.lk; wageesham@uom.lk; pranjeevank@uom.lk).}% <-this % stops a space
\thanks{Corresponding author: J. P. Kulasingham.}}

%\markboth{IEEE Transactions on Audio, Speech, and Language Processing}%
%{Silva \MakeLowercase{\textit{et al.}}: NeuroCodec: EEG-Guided Latent-Domain Generative Model for High-Fidelity Target Speaker 
%Extraction}


\maketitle

\begin{abstract}

Driven by recent advancements in hearing assistive devices, target speaker extraction (TSE) in multi-talker environments has drawn significant attention. To address this, hearing technologies identify attention using neural signals recorded via electroencephalography (EEG) to guide TSE, enabling neuro-steered auditory attention decoding. However, most existing techniques rely on time-frequency masking, which suffers from low fidelity in spectro-temporally overlapping regions and is often restricted to narrowband audio. In contrast, we introduce \textit{NeuroCodec}, which, to the best of our knowledge, is the first generative EEG-guided TSE framework operating in the latent space of a neural audio codec. Specifically, NeuroCodec models spatial EEG dependencies via graph convolutions and fuses cross-modal EEG and speech representations through attention mechanisms and Mamba-based temporal modeling. We further introduce a six-phase training curriculum that progressively incorporates latent reconstruction, multi-scale spectral, and adversarial objectives to ensure stable, high-fidelity speech generation.
Experiments on the public Cocktail Party and KUL datasets
show NeuroCodec outperforms baselines, across several perceptual metrics. A listening experiment  with 20 participants validated that NeuroCodec offers greater perceptual accuracy, achieving a mean score of 82.24 compared with 49.63 for M3ANet.  Additionally, the proposed method is seen to enable 44.1 kHz reconstruction with spectral content extending up to 20 kHz, further supporting the perceptual results. Finally, we extend NeuroCodec with a causal streaming formulation suited for real-time processing, providing a basis for  hearing assistive devices.
\end{abstract}


\vspace{-0.4cm}

\section{Introduction}
\IEEEPARstart{I}{n} complex, multi-talker acoustic environments, the human auditory system exhibits a remarkable capacity to isolate a target speech stream while suppressing interfering voices and background noise, an ability commonly
referred to as selective auditory attention in a “cocktail party” scenario~\cite{bronkhorst2000cocktail, cherry1953some}.
However, this selective auditory attention is severely compromised in listeners with hearing impairments~\cite{shinn2008selective}, leading to degraded speech intelligibility and elevated cognitive load~\cite{pichora2016hearing, peelle2018listening}. While advanced hearing aid technologies and digital acoustic prostheses have successfully restored broad auditory accessibility for millions of individuals, enabling these devices to operate effectively in dynamic, multi-speaker scenes remains a critical hurdle. For modern speech enhancement systems, one fundamental challenge lies not merely in isolating overlapping speech audio, but in decoding the listener's attention to determine \textit{which} speaker to extract from the acoustic mixture.

Target Speaker Extraction (TSE) aims to address this problem by isolating a desired speaker from a speech mixture using auxiliary information about the target speaker~\cite{zmolikova2023neural}. Recent approaches exploit a variety of cues, including acoustic embeddings, directional or binaural spatial cues~\cite {alcalapadilla2025location, indrapala2026bitse}, and visual signals~\cite{wu2024uninet,li2025momuse}, to guide the extraction process and improve robustness in challenging acoustic environments. Although these approaches have achieved strong extraction performance, they rely on external reference information that may
be unavailable or unreliable in dynamic real-world environments.

An alternative approach is to decode the attentional target from the listener's brain activity, which has been shown to be strongly correlated with the attended speech signal. Several studies in the field of auditory neuroscience have demonstrated that the electroencephalography (EEG) signal tracks the attended speech, which is also called neural speech tracking \cite{mesgarani2012selective, ding2012emergence}. Consequently, EEG serves as a direct, non-invasive, and robust indicator of attention, enabling TSE without relying on restrictive external cues. While early computational approaches leveraged this neural-acoustic correlation solely to detect the attended speaker~\cite{biesmans2016auditory} without speech extraction, a task 
known as Auditory Attention Decoding (AAD), recent advances have established integrated neuro-steered TSE frameworks~\cite{fan2025m3anet, pan2024neuroheed, dasilva25_interspeech, 10446333}. By utilizing the continuous EEG signal as the reference cue, these systems directly isolate and reconstruct the attended speech waveform from the multi-talker acoustic mixture~\cite{ceolini2020brain}.

Neuro-steered TSE can be categorized into two main processing paradigms: offline and online extraction. Offline neuro-steered TSE processes complete audio-EEG sequence before producing the target speech, allowing the model to exploit full temporal context during extraction.
Early work,
such as brain-informed speech separation (BISS)~\cite{ceolini2020brain},
used an EEG-decoded speech envelope to guide
TSE. 
The brain-assisted speech enhancement network (BASEN)~\cite{Zhang2023Basen} later introduced convolutional cross-attention for direct EEG-guided speech enhancement, while NeuroHeed~\cite{pan2024neuroheed} employed self-attention-based EEG encoding and an end-to-end extraction framework. NeuroHeed+~\cite{10446333} extended this approach through joint auditory attention detection and extraction. NeuroSpex+~\cite{dasilva25_interspeech} further improves extraction through dual-task training on speech-envelope and waveform objectives and is evaluated under a subject-independent (SI) setting, where test subjects are unseen during training. 
Although SI evaluation is more challenging than subject-dependent (SD) testing, it provides a rigorous measure of cross-subject generalization for practical deployment. 
More recently, M3ANet~\cite{fan2025m3anet} incorporated graph-based EEG modeling and GroupMamba blocks. 

Despite these architectural advances, the inherent reliance on offline processing renders these models unsuitable for real-time deployment in hearing assistive 
devices that require continuous, low-latency operation. To bridge this gap, online processing paradigms handle the input audio stream using short, overlapping 
causal windows. NeuroHeed~\cite{pan2024neuroheed} introduced an online formulation for causal neuro-steered TSE, while CA-NeuroSpex~\cite{de2025neurospex} 
more recently explored a related causal setting. However, causal neuro-steered extraction remains highly challenging, with existing online systems generally showing lower  performance than their offline counterparts \cite{pan2024neuroheed}.

Taken together, prior work leaves two main gaps. First,  existing EEG-guided TSE systems still rely on discriminative mixture-domain mask prediction \cite{fan2025m3anet,pan2024neuroheed}. When target and interfering speech overlap, an inaccurate mask may retain components of the interfering speaker, resulting in speaker leakage. Second, because prior systems have been evaluated at limited audio sampling rates, at a maximum of 8~kHz or 14.7~kHz~\cite{fan2025m3anet, pan2024neuroheed}, their ability to reconstruct wider-band attended speech has not been established.  This narrowband constraint yields an output with a muffled, robotic acoustic quality that fails to meet the perceptual naturalness required for real-world applications.

Generative approaches have recently shown promising performance in speech enhancement and for TSE conditioned on acoustic or visual cues \cite{richter2023speech,navon2025flowtse,wang2024speechx}.
However, these models are computationally intensive, rely on slow iterative generation, and typically benefit from large-scale training data. These requirements are particularly challenging for EEG-guided TSE, where paired EEG-speech datasets are limited and low-latency processing is desirable. To address these constraints, neural audio codecs such as SoundStream~\cite{zeghidour2021soundstream}, EnCodec \cite{defossez2023encodec}, and the Descript Audio Codec (DAC) \cite{kumar2023high}, 
offer an alternative by using Residual Vector Quantization (RVQ) to compress
high-rate audio into compact, low-frame-rate representations that can be reconstructed by a learned decoder. Predicting these target speech latents fundamentally transforms the extraction task from discriminative mixture-domain mask estimation to direct, generative latent synthesis.

In this paper, we propose \textit{NeuroCodec}, a generative neuro-steered TSE framework that addresses the limitations of time-frequency mask estimation using a generative neural audio codec. Specifically, the framework predicts target speech representations within the latent space of a frozen, pretrained DAC \cite{kumar2023high}. This generative formulation aims to reduce masking-related artifacts such as spectral holes and residual speaker leakage, while enabling substantially wider-band reconstruction. Unlike prior neuro-steered systems limited to
an output bandwidth of approximately $4$-$7$~kHz, \textit{NeuroCodec} reconstructs spectral content up to $20$~kHz. Cross-modal attention and Mamba-based temporal modeling fuse EEG and mixture audio representations at the compact codec frame rate.
To the best of our knowledge, there has not been prior work on generative EEG-guided TSE frameworks. Moreover, we present \textit{OnlineNeuroCodec}, a causal streaming extension tailored for real-time hearing assistive devices.

Our main contributions are as follows:
\begin{itemize}
    \item A generative EEG-guided target speech extraction framework, \textit{NeuroCodec}, that predicts target speech in the latent space of a neural audio codec, featuring a six-phase training curriculum to stabilize latent, spectral, and adversarial objectives while mitigating traditional time-frequency masking artifacts.
    \item A causal streaming extension, \textit{OnlineNeuroCodec}, incorporating previous target-latent estimates through an auditory attractor to enable real-time, neuro-steered extraction.
    \item A comprehensive experimental evaluation under the SI setting on the Cocktail Party~\cite{broderick2018electrophysiological} and KUL~\cite{biesmans2016auditory} datasets, featuring objective evaluations, ablation studies, streaming benchmarks, spectrogram analysis, and a perceptual listening study demonstrating reconstruction of spectral content up to approximately 20~kHz.
\end{itemize}

The remainder of the paper is organized as follows. Section~\ref{sec:model} introduces the proposed \textit{NeuroCodec} framework, Section~\ref{sec:online} presents \textit{OnlineNeuroCodec}, the causal extension for online streaming. 
Section~\ref{sec:datasets_and_eval} describes the implementation details and datasets, Section~\ref{sec:results_disc} presents the results and discussion, and Section~\ref{sec:conclusion} is the conclusion.

\begin{figure*}[t]
    \centering
    \includegraphics[width=1\linewidth]{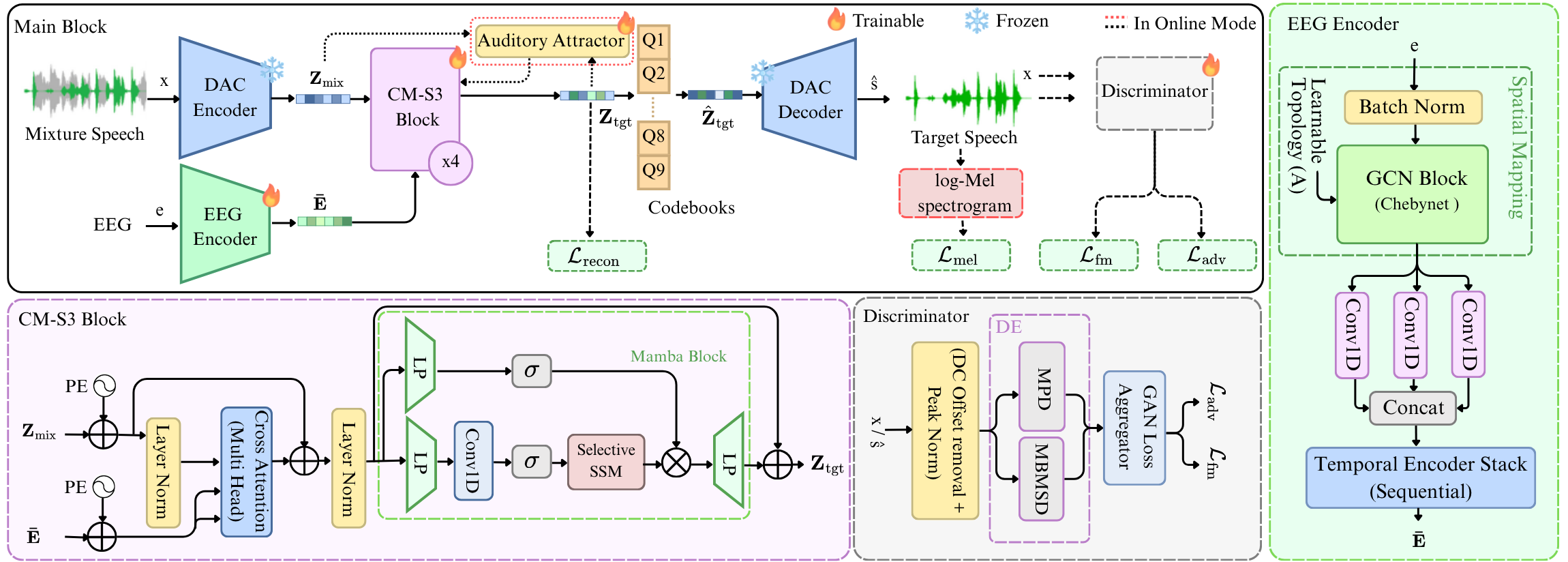}\
    \vspace{-0.5cm}
\caption{Illustration of NeuroCodec architecture. A frozen DAC encoder extracts mixture latents $\mathbf{Z}_{\mathrm{mix}}$, which are fused with EEG embeddings through stacked CM-S3 blocks to predict target latents $\mathbf{Z}_{\mathrm{tgt}}$. The predicted latents are quantized via codebooks ($Q_1$--$Q_9$), and a frozen DAC decoder reconstructs the target speech waveform $\hat{\mathrm{s}}$. The EEG encoder captures spatial-temporal neural features, and multi-period/multi-band discriminators provide adversarial supervision.}
\label{fig:neurocodec_arch}
\vspace{-0.4cm}
\end{figure*}

\vspace{-0.3cm}
\section{NeuroCodec Framework}\label{sec:model}

\subsection{Model Overview}
\vspace{-0.1cm}
% \vspace{-0.2cm}

We formulate the TSE problem as a generative mapping from a multi-speaker mixture and multi-channel EEG signals to the target speaker's latent representations, which are subsequently decoded into a high-fidelity speech waveform. By predicting these latents within the structured space of a frozen pretrained DAC~\cite{kumar2023high}, the framework shifts from mixture-domain interference suppression to cross-modal latent alignment. This enables the model to leverage pretrained perceptual priors for speech extraction while focusing its learning capacity on EEG-guided attention detection. The codec's decoder is utilized for final waveform synthesis. As shown in Fig.~\ref{fig:neurocodec_arch} the architecture consists of four stages: (i) audio encoder, (ii) EEG encoder, (iii) cross-modal selective state space block (CM-S3), and (iv) codec latent generator and decoder. The complete offline training algorithm is provided in the supplementary materials.

\vspace{-0.3cm}
\subsection{Audio Encoder}
\vspace{-0.05cm}

A pretrained and frozen DAC~\cite{kumar2023high} encoder $\mathcal{E}_c$ maps the mixture waveform $\mathbf{x}\in\mathbb{R}^{B\times1\times T}$ to a compact latent
representation $\mathbf{Z}_{\mathrm{mix}}\in\mathbb{R}^{B\times D\times T_c}$, where $B$ is the batch size, $T$ is the number of waveform samples, $T_c$ is the number of latent samples, and $D=1024$ is the codec latent dimension size. The compact codec representation substantially reduces the temporal resolution relative to the raw audio, enabling efficient subsequent EEG-audio fusion. The DAC encoder remains frozen throughout training to preserve its pretrained acoustic representation.

\vspace{-0.3cm}
\subsection{EEG Encoder}
\vspace{-0.05cm}

Adapted from M3ANet~\cite{fan2025m3anet}, the EEG encoder applies
third order Chebyshev graph convolutions~\cite{bai2023sgcrnn} to model
spatial dependencies across electrodes. Unlike prior approaches that
upsample EEG toward the audio rate~\cite{pan2024neuroheed,
fan2025m3anet,dasilva25_interspeech}, NeuroCodec operates at the compact codec
latent rate (50 or 86~Hz), close to the native 128~Hz EEG resolution,
thereby avoiding the initial upsampling and substantially
reducing interpolation and computational overhead. The graph features are
then processed by parallel 1-D convolutions with kernel sizes 3, 7, and 11,
followed by a temporal encoder stack of 1-D convolutions to capture deeper sequential dependencies, that yields a latent output
$\bar{\mathbf{E}}\in\mathbb{R}^{B\times H\times T'_e}$, where $H$ is the EEG embedding dimension and $T'_e$ is the number of latent samples.

\vspace{-0.3cm}
\subsection{Cross Modal Selective State Space Block}
\vspace{-0.05cm}

Positional encodings are then injected into both modalities and the temporal sequences are aligned via multi-head cross-attention~\cite{vaswani2017attention}, defined as     
\vspace{-0.15cm}
\begin{equation}
\mathbf{Z}_1 = \mathbf{Z}_{\mathrm{mix}} + 
\mathrm{Attn}(\mathrm{LN}(\mathbf{Z}_{\mathrm{mix}}), \mathrm{LN}(\mathbf{\bar{E}}), \mathrm{LN}(\mathbf{\bar{E}}))
\vspace{-0.15cm}
\end{equation}
where $\text{LN}(\cdot)$ denotes layer normalization and $\text{Attn}(\cdot, \cdot, \cdot)$  denotes cross-modal attention with inputs of queries, keys and values.  The acoustic tokens act as queries, while the normalized EEG features serve as keys and values to steer the extraction. Long-range temporal dependencies are subsequently modeled using a Mamba block~\cite{gu2024mamba}, defined as 
\vspace{-0.15cm}
\begin{equation}
\mathbf{Z}_{\mathrm{tgt}} = \mathbf{Z}_1 + \mathrm{Mamba}(\mathrm{LN}(\mathbf{Z}_1)) 
\vspace{-0.15cm}
\end{equation}
By replacing self-attention with selective state-space modeling, Mamba provides efficient linear-time temporal modeling of codec-latent sequences.

\vspace{-0.3cm}
\subsection{Codec Latent Generator and Decoder}
\vspace{-0.05cm}

The output of the four stacked CM-S3 blocks is projected back to the codec
latent dimension to yield the continuous target latents
$\tilde{\mathbf Z}_{\mathrm{tgt}} \in \mathbb{R}^{B \times D \times T_c}$. These predicted latents are then quantized using the residual vector quantization (RVQ)~\cite{kumar2023high} codebooks ($Q_1$-$Q_9$) of the frozen DAC, yielding the discrete latent representation defined as $\hat{\mathbf{Z}}_{\mathrm{tgt}} = \mathcal{Q}(\tilde{\mathbf{Z}}_{\mathrm{tgt}})$, where $\mathcal{Q}$ denotes the quantization operation. Finally, a pretrained and frozen DAC decoder $\mathcal{D}_c$ generates the target speech waveform from these quantized latents, providing the target reconstruction defined as $\hat{\mathbf{s}} = \mathcal{D}_c(\hat{\mathbf{Z}}_{\mathrm{tgt}})$.

\vspace{-0.3cm}
\subsection{Loss Functions}
\label{sec:losses}
\vspace{-0.05cm}

Unlike conventional models that optimize scale-invariant signal-to-distortion ratio (SI-SDR) loss directly on the decoded audio \cite{ceolini2020brain, pan2024neuroheed, fan2025m3anet, dasilva25_interspeech}, Neurocodec is trained using a composite objective function for accurate latent alignment and high-fidelity acoustic reconstruction.

Reconstruction loss, $\mathcal{L}_{\mathrm{recon}}$, is computed as the mean squared error between the predicted and ground-truth target continuous latents and is used to guide the model to reconstruct the target speaker representation in the codec latent space.

However, reconstruction loss alone can yield perceptual artifacts. Spectral loss, $\mathcal{L}_{\mathrm{mel}}$~\cite{kumar2023high,guo2022multi}, improves spectral fidelity by minimizing the $L_1$ distance between log-Mel spectrograms at multiple resolutions. We use seven scales up to 20~kHz for the Cocktail Party~\cite{broderick2018electrophysiological} dataset and four scales up to 4~kHz for KUL dataset~\cite{biesmans2016auditory} to capture both transient and steady-state spectral characteristics across their respective bandwidths.

To further improve perceptual realism, we use an adversarial ensemble of Multi-Period Discriminators
(MPD)~\cite{kong2020hifi} to model waveform periodicity and a Multi-band Multi-scale Short-Time Fourier Transform (STFT) Discriminator (MBMSD)~\cite{kumar2023high} to enforce multi-band spectral consistency. The generator is optimized using
hinge adversarial loss $\mathcal{L}_{\mathrm{adv}}$ and $L_1$ feature-matching loss
$\mathcal{L}_{\mathrm{fm}}$ on intermediate discriminator features, while the
discriminators are trained with the standard hinge objective.

\vspace{-0.3cm}
\subsection{Multi-phase Training Curriculum }
\vspace{-0.05cm}

The overall generator objective is given by,
\begin{equation}
\mathcal{L}_{G} =
\lambda_{\mathrm{recon}}\mathcal{L}_{\mathrm{recon}} +
\lambda_{\mathrm{mel}}\mathcal{L}_{\mathrm{mel}} +
\lambda_{\mathrm{adv}}\mathcal{L}_{\mathrm{adv}} +
\lambda_{\mathrm{fm}}\mathcal{L}_{\mathrm{fm}},
\end{equation}
where $\lambda_{\mathrm{recon}}$, $\lambda_{\mathrm{mel}}$, $\lambda_{\mathrm{adv}}$, and $\lambda_{\mathrm{fm}}$
are weighting coefficients balancing their respective loss terms throughout training. We set $\lambda_{\mathrm{recon}}=1$ to be fixed throughout training.

To preserve GAN stability and mitigate mode collapse during training, we employ a six-phase curriculum that progressively introduces spectral and adversarial objectives. In Phase~1, the model optimizes exclusively $\mathcal{L}_{\mathrm{recon}}$ until the validation latent mean squared error (MSE)
falls below 0.87, establishing a stable latent mapping prior to multi-modal supervision. Phase 2 gradually introduces the Mel loss ($\mathcal{L}_{\mathrm{mel}}$),
ramping $\lambda_{\mathrm{mel}}$ from 2 to 15 to progressively incorporate spectral supervision without destabilizing the learned latent representation. Upon reaching 
$\lambda_{\mathrm{mel}} = 15$, Phase~3 maintains joint latent and spectral 
optimization until the validation loss decreases for at least one epoch. In Phase~4, the generator parameters are frozen for 2 epochs to allow warm-up pre-training of the 
discriminators, providing stable adversarial supervision. Phase~5 then
jointly optimizes the generator and discriminators, progressively increasing $(\lambda_{\mathrm{adv}},\lambda_{\mathrm{fm}})$ in steps of $(0.25,0.5)$ 
until the target values of $(1.0,2.0)$ are reached. Finally, Phase~6 
conducts full adversarial training until validation performance plateaus, employing a
\texttt{ReduceLROnPlateau} scheduler with patience of 3 epochs and a reduction factor of 0.5. 

In the following section, we extend this framework into a causal streaming formulation, \textit{OnlineNeuroCodec}, which integrates an auditory attractor mechanism to enable real-time target speech extraction.

\vspace{-0.3cm}
\section{Causal streaming extension of NeuroCodec} \label{sec:online}

We extend NeuroCodec to a causal formulation, termed \textit{OnlineNeuroCodec}. The causal model preserves the latent-domain processing of the offline formulation while restricting each prediction to the current input window and the information accumulated from preceding windows.

Inference uses a sliding window comprising buffered context $w_b$ and a
new input hop $w_c$, following NeuroHeed~\cite{pan2024neuroheed}.
At each hop, only the latent frames corresponding to $w_c$ are generated,
while the preceding $w_b$ frames provide causal context. The complete streaming procedure, including the cold start and the running speaker state introduced below, 
is given in Algorithm~\ref{alg:online}.

\begin{algorithm}[t]
 \caption{OnlineNeuroCodec: causal streaming inference.}
 \label{alg:online}
 \begin{algorithmic}[1]

 \REQUIRE Streaming mixture $\mathbf{x}$, EEG $\mathbf{E}$;
 window $w_b$, hop $w_c$
 \REQUIRE Frozen $\mathcal{E}_c$,  $\mathcal{Q}$,
 and  $\mathcal{D}_c$; EMA rate $\alpha$
 \ENSURE Extracted waveform $\hat{\mathbf{s}}$

 \STATE $\mathbf{Z}_{\mathrm{past}}\leftarrow\emptyset$;
 $\hat{\mathbf{s}}\leftarrow\emptyset$

 \FOR{each hop $h=1,\dots,N_{\mathrm{hops}}$}

     \STATE $\mathbf{x}^{(h)},\mathbf{R}^{(h)}
     \leftarrow$ window of length $w$ ending at hop $h$

     \STATE $\mathbf{Z}_{\mathrm{mix}}^{(h)}
     \leftarrow\mathcal{E}_c(\mathbf{x}^{(h)})$;
     $\bar{\mathbf{E}}^{(h)}
     \leftarrow\mathrm{EEGEnc}(\mathbf{R}^{(h)})$

     \IF{$h=1$}
         \STATE $\mathbf{Z}_{\mathrm{pseudo}}^{(h)}
         \leftarrow
        \mathrm{CM\!-\!S3}
         (\mathbf{Z}_{\mathrm{mix}}^{(h)},\bar{\mathbf{E}}^{(h)})$
         \STATE $\mathbf{Z}_{\mathrm{past}}
         \leftarrow\mathbf{Z}_{\mathrm{pseudo}}^{(h)}$
     \ENDIF

     \STATE $[\gamma,\beta]
     \leftarrow
     \mathrm{Linear}
     \!\left(\mathrm{Mean}(\mathbf{Z}_{\mathrm{past}})\right)$

     \STATE $\widetilde{\mathbf{Z}}_{\mathrm{mix}}^{(h)}
     \leftarrow
     (1+\gamma)\odot
     \mathrm{LN}(\mathbf{Z}_{\mathrm{mix}}^{(h)})+\beta$

     \STATE $\widetilde{\mathbf{Z}}_{\mathrm{tgt}}^{(h)}
     \leftarrow
     \mathrm{CM\!-\!S3}
     (\widetilde{\mathbf{Z}}_{\mathrm{mix}}^{(h)},\bar{\mathbf{E}}^{(h)})$

     \STATE $\hat{\mathbf{Z}}_{\mathrm{tgt}}^{(h)}
     \leftarrow
     \mathcal{Q}(\widetilde{\mathbf{Z}}_{\mathrm{tgt}}^{(h)})$

     \STATE $\mathbf{Z}_{\mathrm{past}}
     \leftarrow
     \alpha\mathbf{Z}_{\mathrm{past}}
     +(1-\alpha)\hat{\mathbf{Z}}_{\mathrm{tgt}}^{(h)}$

     \STATE $\mathbf{Z}_{\mathrm{past}}
     \leftarrow
     \mathrm{detach}(\mathbf{Z}_{\mathrm{past}})$

     \STATE $\hat{\mathbf{s}}
     \leftarrow
     \hat{\mathbf{s}}\Vert
     \mathcal{D}_c(\hat{\mathbf{Z}}_{\mathrm{tgt}}^{(h)})$

 \ENDFOR

 \RETURN $\hat{\mathbf{s}}$

 \end{algorithmic}

 \end{algorithm}
 
\vspace{-0.3cm}
\subsection{Auditory Attractor via Running Speaker State}
\vspace{-0.05cm}

OnlineNeuroCodec preserves attended-speaker identity across hops by
maintaining a running latent representation of previous target estimates,
$\mathbf{Z}_{\mathrm{past}}$. From the second hop onward, it is updated using an
Exponential Moving Average (EMA)~\cite{tarvainen2017mean}:
\vspace{-0.2cm}
\begin{equation}
\mathbf{Z}_{\mathrm{past}}
\leftarrow
\alpha\mathbf{Z}_{\mathrm{past}}
+(1-\alpha)\hat{\mathbf{Z}}^{\,h}_{\mathrm{tgt}},
\label{eq:ema}
\vspace{-0.15cm}
\end{equation}

\noindent where $\hat{\mathbf{Z}}^{\,h}_{\mathrm{tgt}}$ denotes the target-latent estimate of the current hop and $\alpha$ is the EMA smoothing coefficient, set to 0.7 based on validation performance. The EMA combines accumulated speaker information with the latest estimate, providing a stable speaker memory for subsequent extraction.

OnlineNeuroCodec therefore uses two complementary cues: a \emph{neuronal attractor}, in which EEG features provide the keys and values for cross-attention, and an \emph{auditory attractor} derived from $\mathbf{Z}_{\mathrm{past}}$. The latter is mean-pooled and linearly projected to FiLM~\cite{perez2017visual} scale and shift parameters $(\gamma,\beta)$,
which modulate the current mixture-latent query $\mathbf{Z}_{\mathrm{mix}}$ before EEG cross-attention:
\vspace{-0.15cm}
\begin{equation}
[\gamma,\beta]
=
\mathrm{Linear}\!\left(\mathrm{Mean}(\mathbf{Z}_{\mathrm{past}})\right),
\label{eq:film_params}
\end{equation}
\vspace{-1cm}

\begin{equation}
\widetilde{\mathbf{Z}}_{\mathrm{mix}}
=
(1+\gamma)\odot\mathrm{LN}(\mathbf{Z}_{\mathrm{mix}})+\beta .
\label{eq:film}
\vspace{-0.1cm}
\end{equation}

\noindent Thus, previous target estimates provide speaker-dependent conditioning for the current hop. Similar to the autoregressive attractor in NeuroHeed~\cite{pan2024neuroheed}, this mechanism self-enrolls the attended speaker over time. However, NeuroHeed re-encodes previous waveform outputs using a dedicated LSTM speaker encoder, whereas OnlineNeuroCodec
updates the speaker state directly in codec latent space using a parameter-free EMA, avoiding an additional recurrent speaker module.

\vspace{-0.3cm}
\subsection{Hopwise Training Strategy}
\vspace{-0.05cm}

At the first hop, no preceding prediction is available. Therefore, $\mathbf Z_{\mathrm{past}}$ is initialized from an EEG-conditioned forward pass without auditory-history conditioning. During training, speaker dropout with probability $p=0.2$ is applied so that the model remains functional when the historical context is unavailable \cite{bengio2015scheduled}. 

Training follows the streaming setting by applying the reconstruction and spectral objectives of Section~\ref{sec:losses} to each output hop. For hop $h$, $\mathcal{L}_{\mathrm{recon}}^{(h)}$ operates in the latent domain, whereas $\mathcal{L}_{\mathrm{mel}}^{(h)}$ is evaluated on the corresponding decoded speech segment. The overall online objective is
\vspace{-0.15cm}
\begin{equation}
\mathcal{L}_{\mathrm{online}}
=
\frac{1}{N_{\mathrm{hops}}}
\sum_{h=1}^{N_{\mathrm{hops}}}
\left(
\lambda_{\mathrm{recon}}\mathcal{L}_{\mathrm{recon}}^{(h)}
+
\lambda_{\mathrm{mel}}\mathcal{L}_{\mathrm{mel}}^{(h)}
\right),
\label{eq:online_loss}
\vspace{-0.1cm}
\end{equation}

\noindent where $N_{\mathrm{hops}}$ is the number of output hops, and $\lambda_{\mathrm{recon}}$ and $\lambda_{\mathrm{mel}}$ weight the reconstruction and spectral terms, respectively. Averaging across hops keeps the objective scale independent of sequence length.

During training, gradients are stopped at $\mathbf{Z}_{\mathrm{past}}$ between consecutive hops, truncating backpropagation across hop boundaries while maintaining the state for causal conditioning.

\vspace{-0.3cm}
\section{Datasets and Experiments}\label{sec:datasets_and_eval}

In this section, we detail the experimental setup and the datasets used to evaluate the proposed method.

\vspace{-0.3cm}
\subsection{Datasets}
\vspace{-0.05cm}

To evaluate our proposed method, we used two publicly available multi-speaker EEG-audio datasets: Cocktail Party and KUL. The Cocktail Party dataset~\cite{broderick2018electrophysiological} includes EEG recordings from 33 subjects (128 channels, 512~Hz) who completed 30 dichotic listening trials (60~s each), attending to one of two simultaneous narrative speech streams presented to opposite ears. The KUL data set~\cite{biesmans2016auditory} consists of 16 subjects with normal-hearing with 20 trials each; in our experiments, we used the first eight trials (similar to prior work for fair comparison~\cite{pan2024neuroheed,dasilva25_interspeech}), such that subjects do not attend to the same speech stimulus repeatedly, where participants selectively attended one of two concurrent Dutch narrative stories presented dichotically.

\vspace{-0.3cm}
\subsection{Implementation Details}
\vspace{-0.05cm}

EEG signals are sampled at 128~Hz and were band-pass filtered between 0.1-45~Hz, followed by ICA-based artifact removal~\cite{delorme2004eeglab}. All experiments use subject independent (SI) evaluation, where test subjects
are unseen during training, providing a more stringent assessment of cross-subject generalization.

For the Cocktail Party dataset, EEG is additionally re-referenced to the average mastoids. Subjects are randomly split with a fixed seed into 23/5/5, training/validation/test subjects, and the same SI split is used for all compared models. For fair comparison with M3ANet~\cite{fan2025m3anet}, BASEN~\cite{Zhang2023Basen}, and MSFNet~\cite{fan2024msfnet}, trials are segmented into 2s non-overlapping windows. The native 44.1~kHz audio is processed using the pretrained 44.1~kHz DAC encoder~\cite{kumar2023high}, which has a downsampling factor of 512 and yields an approximately 86~Hz latent frame rate.

For KUL, we use the preprocessed data released with NeuroHeed~\cite{pan2024neuroheed} and perform 16-fold SI cross-validation. For fair comparison with the SI NeuroSpex+ setting~\cite{dasilva25_interspeech}, recordings are segmented into 4s windows with a 1-s hop. The original 8~kHz audio is upsampled to 16~kHz and processed using the pretrained 16~kHz DAC encoder~\cite{kumar2023high}, which has a downsampling factor of 320 and yields a 50~Hz latent frame rate. Note that the difference in window sizes between the two datasets is to ensure fair comparison with the prior work on each dataset. Additional implementation details and hyperparameters are provided in the supplementary material. The source code will be released upon acceptance.

\vspace{-0.3cm}
\subsection{Evaluation Metrics}
\vspace{-0.05cm}

Performance is evaluated using standard objective speech enhancement metrics: Short-Time Objective Intelligibility (STOI)~\cite{taal2011algorithm}, Extended Short-Time Objective Intelligibility (ESTOI)~\cite{7539284}, and Perceptual Evaluation of Speech Quality (PESQ)~\cite{rix01_icassp}. To ensure fair comparison with existing baselines that only operate at 8~kHz sample rate, all outputs are resampled to 8~kHz before objective evaluation. This prevents the wider output bandwidth of NeuroCodec from providing an inherent advantage in the reported objective metrics.

\vspace{-0.3cm}
\section{Results and discussion}\label{sec:results_disc}
This section details the experimental evaluation of the proposed framework across two public datasets. We present (i) objective speech quality metrics, (ii) an ablation study validating our core architectural choices, (iii) a subjective listening test, and (iv) a performance evaluation of the causal extension.

\vspace{-0.3cm}
\subsection{Experimental Results} \label{sec:results}
\vspace{-0.05cm}
Table~\ref{tab:combined_results} compares NeuroCodec with previous EEG-guided target speaker extraction systems on the Cocktail Party and KUL datasets. 

On the Cocktail Party dataset, BASEN~\cite{Zhang2023Basen} and M3ANet~\cite{fan2025m3anet} are reproduced using the authors' publicly available implementations to ensure a fair comparison by evaluating under the same SI split used for NeuroCodec. NeuroCodec achieves the highest STOI and PESQ values, improving STOI from 0.812 (M3ANet) to 0.827, and PESQ from 2.23 to 2.31. While M3ANet retains a marginal advantage in ESTOI, these results demonstrate that NeuroCodec improves overall speech intelligibility and perceptual quality, while remaining highly competitive in ESTOI.

On the KUL dataset, the performance metrics for prior baselines BISS~\cite{ceolini2020brain}, NeuroHeed~\cite{pan2024neuroheed}, and Neuro-Spex~\cite{de2024neurospex} are sourced directly from the Neuro-Spex+ paper~\cite{dasilva25_interspeech}, which reports the identical SI configuration. The BISS baseline shows substantially lower performance compared to all other models, as it employs a subject-dependent linear decoder for envelope reconstruction that is trained exclusively on single-talker data. In contrast, when compared with the state-of-the-art NeuroSpex+, our NeuroCodec improves STOI from 0.859 to 0.868 and PESQ from 2.514 to 2.562.

Note that other relevant models such as NeuroHeed+~\cite{10446333} and MSFNet~\cite{fan2024msfnet} are omitted from Table~\ref{tab:combined_results} because the lack of publicly available code prevents re-evaluation under the same rigorous SI setting.

\begin{table}[h!]
\centering
\renewcommand{\arraystretch}{1.1} 
\caption{STOI, ESTOI, and PESQ results on Cocktail Party and KUL}

\label{tab:combined_results}
\setlength{\tabcolsep}{7pt}
\begin{tabularx}{\columnwidth}{X ccc}
\toprule
\textbf{Model} & \textbf{STOI} & \textbf{ESTOI} & \textbf{PESQ}  \\
\midrule
\textit{(a) Cocktail Party Dataset} & & &\\
% NeuroHeed~\cite{pan2024neuroheed}  & 0.715 & 0.548 & 1.45  \\
BASEN~\cite{Zhang2023Basen}     & 0.790 & 0.715 & 1.91   \\
M3ANet~\cite{fan2025m3anet}    & 0.812 & \textbf{0.725} & 2.23  \\
\text{NeuroCodec} & \textbf{0.8273} & 0.721 & \textbf{2.31}  \\
\midrule
\textit{(b) KUL Dataset} & & & \\
BISS~\cite{ceolini2020brain}  & 0.106    & --    & -0.045 \\
NeuroHeed~\cite{pan2024neuroheed}  & 0.637    & --    & 2.282 \\
NeuroSpex~\cite{de2024neurospex} & 0.795 & --    & 2.319  \\
NeuroSpex+~\cite{dasilva25_interspeech} & 0.859 & --    & 2.514 \\
NeuroCodec & \textbf{0.868} & \textbf{0.754} & \textbf{2.562}  \\

\bottomrule
\end{tabularx}
\\ \vspace{6pt} 
{\raggedright \footnotesize \noindent Scores for the Cocktail Party dataset were produced using publicly available code from the relevant authors. Scores for prior models on the KUL dataset are sourced from NeuroSpex+ \cite{dasilva25_interspeech}. Baselines differ across datasets according to their reported evaluation settings. \par} 
\end{table}

\begin{figure}[t]
    \centering
    \includegraphics[width=1\linewidth]{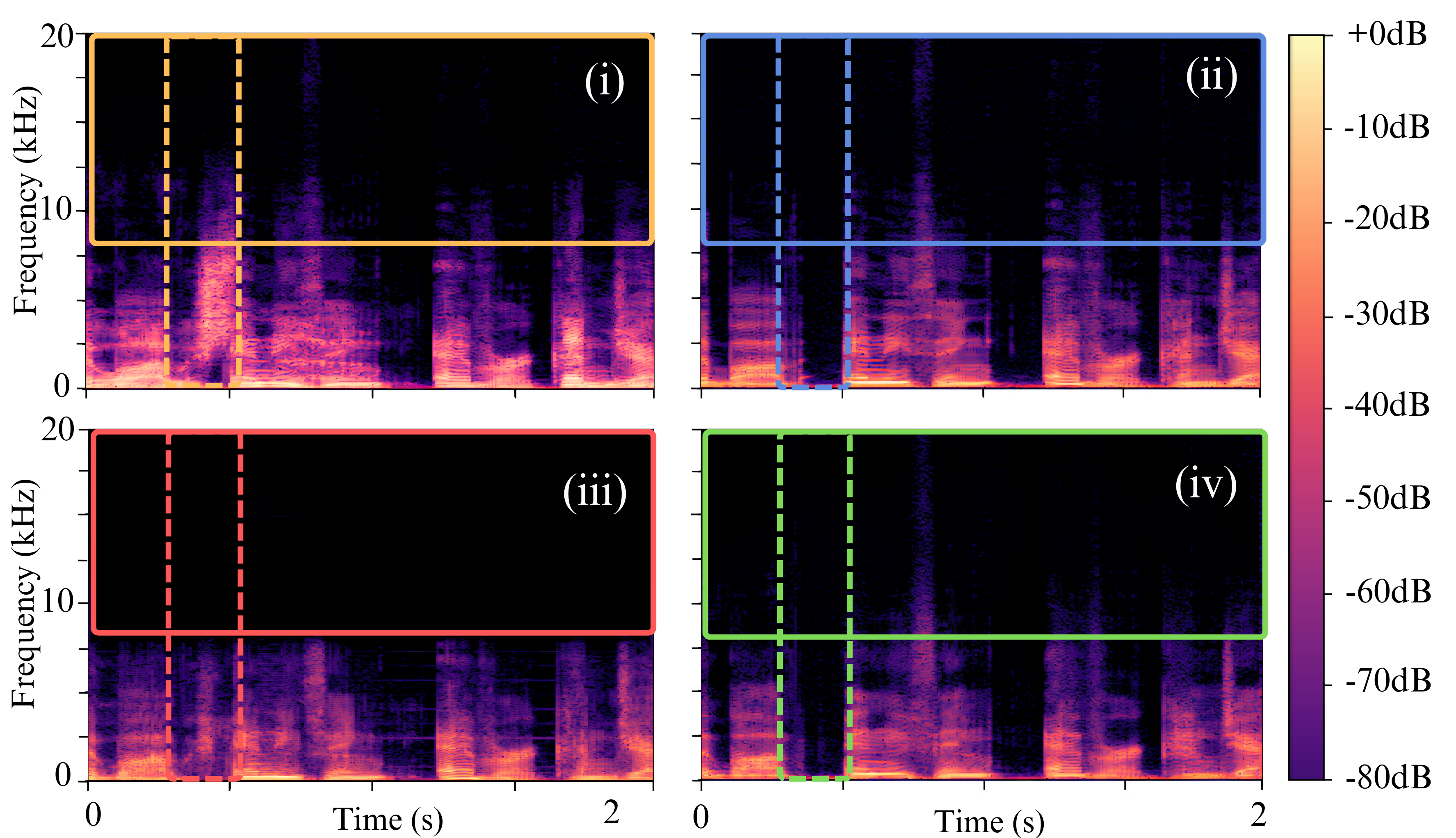}\
    \vspace{-0.5cm}
\caption{Qualitative spectrogram comparison on the Cocktail Party dataset: (i) mixture, (ii) ground-truth target, (iii) M3ANet output, and (iv) NeuroCodec output. Solid boxes highlight NeuroCodec's preserved content above 7~kHz, while dashed region mark overlapping-speech regions with reduced interference. Note that M3ANet does not preserve content above 7 kHz, nor does it avoid speaker leakage within the dashed region.}
\label{fig:Qualitative_Comparision}
\vspace{-0.4cm}
\end{figure}

Qualitative evaluation of the representative spectrograms shown in Fig.~\ref{fig:Qualitative_Comparision} further illustrates these differences. M3ANet~\cite{fan2025m3anet} exhibits high-frequency truncation around 7~kHz, whereas NeuroCodec reconstructs spectral content up to 20~kHz with clearer harmonic structure. In overlapping-speech regions, NeuroCodec also shows cleaner target-speaker separation, consistent with its direct generation of target speech latents rather than masking the mixture representation.

\vspace{-0.3cm}
\subsection{Ablation Study}
\vspace{-0.05cm}

To understand the contribution of each architectural component, we conduct controlled ablation experiments on the KUL dataset. As summarized in Table~\ref{tab:merged_ablation}, we evaluate the proposed architecture across three core dimensions: the fusion block backbone, the internal EEG encoder architecture, and the training objectives. 

\subsubsection{Temporal modeling architecture}
We first investigate the temporal modeling architecture by comparing three sequence modeling strategies: (i) a standard Transformer~\cite{vaswani2017attention}, (ii) a Transformer with Snake activations~\cite{li2023snakegan}, and (iii) the proposed Causal Mamba. The baseline Transformer achieves lower scores, while the Snake-enhanced Transformer improves all three metrics over the standard Transformer. However, the Causal Mamba achieves the highest scores across all three metrics. 

\subsubsection{EEG encoder guidance and components}
Next, we validate the necessity and design of the EEG encoder. To assess the model's dependence on EEG guidance, we replace the EEG input with Gaussian noise. The substantial performance drop supports the conclusion that the model relies on neural guidance for target extraction. Furthermore, ablating specific internal components of the EEG encoder, namely the GCN spatial mapping and the Multi-Scale 1D Convolutions (MS-Conv1D), confirmed that capturing both cross-channel spatial topologies and multi-scale temporal receptive fields is crucial to extract better EEG features. 

\subsubsection{Progressive training objectives}
Finally, we compare the progressive addition of training objectives. Using solely the latent-domain reconstruction loss ($\mathcal{L}_{\mathrm{recon}}$) serves as our baseline. Integrating the multi-scale log-Mel spectrogram loss ($\mathcal{L}_{\mathrm{mel}}$) drastically improves perceptual quality, while the further addition of the adversarial feature-matching loss ($\mathcal{L}_{\mathrm{adv/fm}}$) leads to even more improvement. The combined objective ($\mathcal{L}_{\mathrm{recon}} + \mathcal{L}_{\mathrm{mel}} + \mathcal{L}_{\mathrm{adv/fm}}$) achieves the highest overall performance, reaching a PESQ of 2.5623 and an ESTOI of 0.7544.

\begin{table}[t]
\centering
\renewcommand{\arraystretch}{1.1} 
\caption{Ablation studies of NeuroCodec on the KUL dataset}
\label{tab:merged_ablation}
\resizebox{\columnwidth}{!}{%
\begin{tabular}{lccc}
\toprule
\textbf{Configuration} & \textbf{STOI} & \textbf{ESTOI} & \textbf{PESQ} \\
\midrule
\multicolumn{4}{l}{\textit{(a) Temporal Modeling Architecture (with full EEG Encoder \& $\mathcal{L}_{recon}$)}} \vspace{1 pt}\\ 
\quad Transformer & 0.6120 & 0.4703 & 0.8702 \\
\quad Transformer + Snake & 0.6480 & 0.5162 & 1.0340 \\
\quad \textbf{Mamba} & \textbf{0.7112} & \textbf{0.5734} & \textbf{1.5030} \\
\midrule
\multicolumn{4}{l}{\textit{(b) EEG Encoder Architecture (with Mamba \& $\mathcal{L}_{recon}$)}} \vspace{1 pt}\\
\quad w/o EEG guidance (noise instead of EEG) & 0.5412 & 0.3921 & 0.7845 \\
\quad w/o spatial mapping (w/o GCN) & 0.6725 & 0.5312 & 1.2840 \\
\quad w/o Multi-scale conv1D & 0.6854 & 0.5408 & 1.3912 \\
\quad \textbf{Full Proposed EEG Encoder} & \textbf{0.7112} & \textbf{0.5734} & \textbf{1.5030} \\
\midrule
\multicolumn{4}{l}{\textit{(c) Training Objectives (with Mamba \& full EEG Encoder)}} \vspace{1 pt}\\
\quad $\mathcal{L}_{\mathrm{recon}}$ (Base) & 0.7112 & 0.5734 & 1.5030 \\
\quad $\mathcal{L}_{\mathrm{recon}} + \mathcal{L}_{\mathrm{mel}}$ & 0.8015 & 0.6861 & 1.9204 \\
\quad $\mathcal{L}_{\mathrm{recon}} + \mathcal{L}_{\mathrm{adv/fm}}$ & 0.8242 & 0.7122 & 2.1032 \\
\quad \textbf{$\mathcal{L}_{\mathrm{recon}} + \mathcal{L}_{\mathrm{mel}} + \mathcal{L}_{\mathrm{adv/fm}}$ (Final)} & \textbf{0.8681} & \textbf{0.7544} & \textbf{2.5623} \\
\bottomrule
\end{tabular}%
}
\end{table}

\vspace{-0.3cm}
\subsection{Perceptual Evaluation}
\label{sec:perceptual}
\vspace{-0.05cm}

To verify the perceptual quality of the generated output, we conducted a controlled multi-stimulus listening test with 20 participants. The study was approved by the Ethics Review Committee of the University of Moratuwa (No EDN/2025/012) and all participants provided written informed consent. We used web Multiple Stimulus with Hidden Reference and Anchor (MUSHRA) ~\cite{schoeffler2018webmushra} to compare NeuroCodec's native 44.1kHz output with the native output of the strongest reproducible baseline, M3ANet \cite{fan2025m3anet}. The Cocktail Party dataset was used; KUL was excluded because NeuroSpex+ \cite{dasilva25_interspeech} has no public implementation. Eight 2s samples were randomly selected from the held-out test set.

\begin{figure}[h]
	\centering
	\includegraphics[width=1\linewidth]{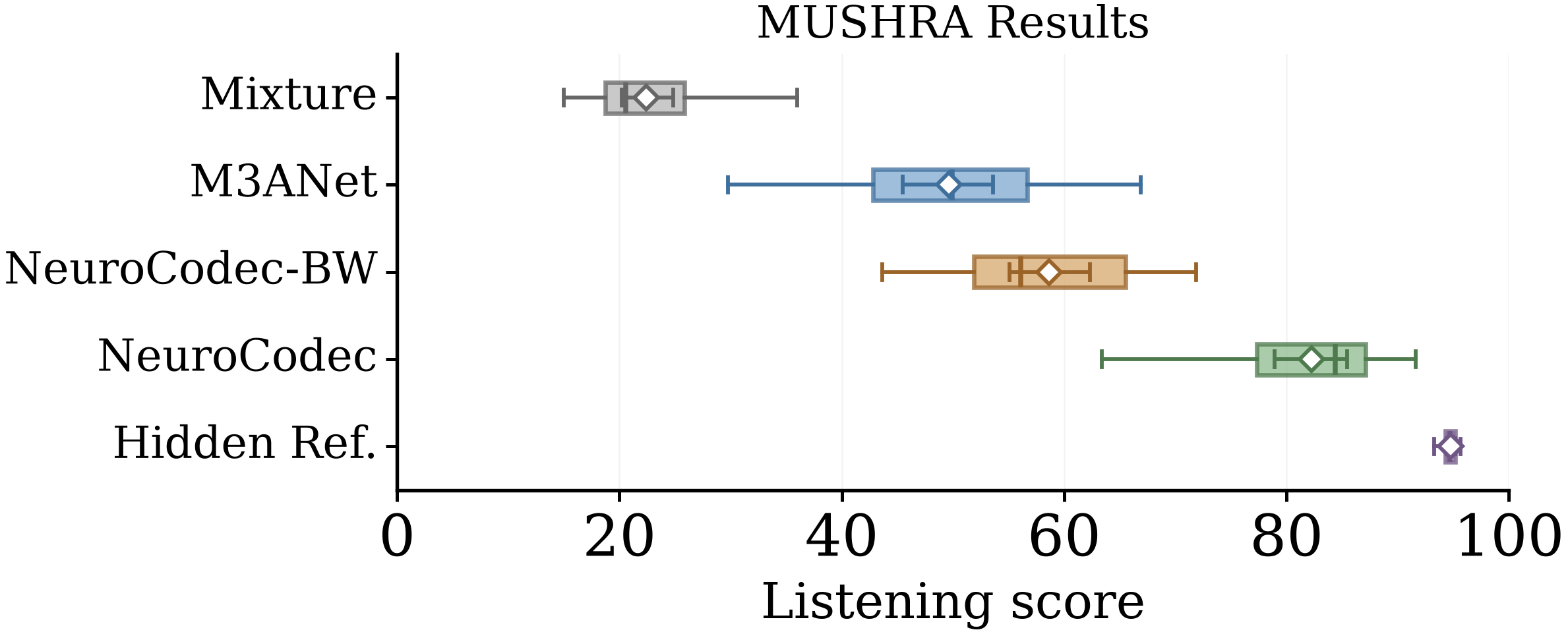}\
	\vspace{-0.5cm}
	\caption{MUSHRA listening-test results on the Cocktail Party dataset. Bandwidth-matched NeuroCodec outperforms M3ANet, while native NeuroCodec achieves a substantially larger perceptual gain.}
	\label{fig:mushra_boxplot}
\end{figure}

Each trial presented five signals under blinded and randomized conditions: the hidden clean reference, mixture, M3ANet, bandwidth-matched NeuroCodec, and NeuroCodec under blinded and randomized presentation. As shown in Fig.~\ref{fig:mushra_boxplot}, the hidden clean reference achieved a mean score of 94.76 (95\% CI: 94.41-95.14) while  NeuroCodec achieved 82.24 (CI: 78.86-85.39).Bandwidth-matched NeuroCodec, obtained by low-pass filtering the NeuroCodec output at 7.296~kHz to match the effective bandwidth of M3ANet, achieved 58.49 (CI: 55.06-62.31), compared with 49.63 (CI: 45.42-53.71) for M3ANet. This improvement under matched bandwidth indicates that NeuroCodec's perceptual advantage is not solely attributable to its wider output bandwidth and is consistent with benefits from the proposed extraction architecture. The substantially higher score of the original NeuroCodec output further demonstrates the perceptual benefit of wider-band speech reconstruction. The mixture obtained a mean score of 22.41 (CI: 20.24-24.80). The NeuroCodec-M3ANet difference was significant under a paired permutation test ($p=3.0\times10^{-5}$).

\vspace{-0.3cm}
\subsection{OnlineNeuroCodec Evaluation}
\vspace{-0.05cm}

In addition to the offline evaluation, we assess OnlineNeuroCodec against our reimplementation of online NeuroHeed~\cite{pan2024neuroheed}.\footnote{The public NeuroHeed repository provides only the offline implementation. CA-NeuroSpex~\cite{de2025neurospex} is omitted because no public implementation is available for reproduction.} OnlineNeuroCodec and the reimplemented NeuroHeed are both evaluated under an identical streaming configuration with a window length of $w = 2.0$~s and a hop size of $w_c = 0.5$~s, with all benchmarks measured on a single GTX 1080 Ti GPU. Computational efficiency is evaluated using the Real-Time Factor ($\text{RTF} = T_{\mathrm{proc}}/w_c$), where $T_{\mathrm{proc}}$ is the processing time. We note that  NeuroHeed ~\cite{pan2024neuroheed} defines RTF using the reciprocal convention, $w_c/T_{\mathrm{proc}}$.

As shown in Table~\ref{tab:online_comparision}, OnlineNeuroCodec attains substantially higher extraction accuracy. Relative to NeuroHeed~\cite{pan2024neuroheed}, STOI improves from 0.612 to 0.803, ESTOI from 0.5716 to 0.7105, and PESQ from 1.290 to 1.662. 

\begin{table}[h!]
\centering
\renewcommand{\arraystretch}{1.1} 
\caption{Online evaluation of OnlineNeuroCodec against NeuroHeed~\cite{pan2024neuroheed} on the KUL dataset.}

\label{tab:online_comparision}
\setlength{\tabcolsep}{3pt}
\begin{tabularx}{\columnwidth}{X cccc}
\toprule
\textbf{Model} & \textbf{STOI$\uparrow$} & \textbf{ESTOI$\uparrow$} & \textbf{PESQ$\uparrow$} & \textbf{RTF$\downarrow$} \\
\midrule
NeuroHeed~\cite{pan2024neuroheed} & 0.588 & 0.497 & 1.037 & 0.27 \\
\begin{tabular}[c]{@{}l@{}}NeuroHeed \cite{pan2024neuroheed} \\ \, \textit{(with autoregressive Speaker Encoder)}\end{tabular} & 0.612 & 0.572 & 1.290 & 0.41 \\
OnlineNeuroCodec & 0.742 & 0.686 & 1.501 & \textbf{0.13} \\
\begin{tabular}[c]{@{}l@{}}OnlineNeuroCodec \\ \, \textit{(with Auditory Attractor)}\end{tabular} & \textbf{0.803} & \textbf{0.711} & \textbf{1.662} & 0.18 \\
\bottomrule
\end{tabularx}
\\ \vspace{6pt} 
{\footnotesize \noindent \raggedright  NeuroHeed results are from our reimplementation, as the authors release code for the offline model only. \par}
\end{table}

Enabling the auditory attractor improves OnlineNeuroCodec from 0.742 to 0.803 STOI and from 1.501 to 1.662 PESQ. Since $\mathbf{Z}_{\mathrm{past}}$ affects the model only in mean-pooled form (Eq.~\ref{eq:film}), this improvement is consistent with more stable tracking of the attended speaker across hops. A comparable trend is observed for NeuroHeed, whose autoregressive speaker encoder raises STOI from 0.588 to 0.612, consistent with the self-enrollment behaviour reported in their work~\cite{pan2024neuroheed}.

OnlineNeuroCodec without the attractor achieves the lowest RTF (0.13). Enabling the auditory attractor increases RTF modestly from 0.13 to 0.18. NeuroHeed is the most expensive of the four configurations (0.27-0.41). OnlineNeuroCodec is therefore approximately 2.1$\times$ faster than NeuroHeed without the attractor mechanism (0.13 against 0.27) and 2.3$\times$ faster with it (0.18 against 0.41), so its accuracy advantage is obtained at approximately half the measured RTF, despite operating on a generative rather than a masking objective.

Table~\ref{tab:online_window} reports the effect of the two streaming parameters. Lengthening the buffered context $w_b$ improves PESQ, since a longer causal history is available for cross-modal alignment, but raises the RTF because each hop must reprocess a longer window. The gain in PESQ saturates beyond $w_b = 1.5$~s while the RTF continues to grow. Enlarging the hop $w_c$ lowers the RTF, as fewer forward passes are required per second of emitted audio, but increases algorithmic latency because the system must wait for a full hop of new audio before it can respond. Hence, the configuration $w_b = 1.5$~s and $w_c = 0.5$~s is adopted throughout, as it attains near-maximal PESQ while maintaining a low RTF.
\vspace{-0.3cm}

\begin{table}[h!]
\centering
\caption{Effect of the buffered context size $w_b$ and the hop size $w_c$ on OnlineNeuroCodec.}
\label{tab:online_window}
\setlength{\tabcolsep}{5pt}
\begin{tabularx}{\columnwidth}{X ccc}
\toprule
 & \textbf{$w$ (s)} & \textbf{PESQ~$\uparrow$} & \textbf{RTF~$\downarrow$} \\
\midrule
\textit{(a) Buffered context $w_b$ ($w_c = 0.5$~s)} & & & \\
\quad $w_b = 0.5$~s & 1.0 & 1.412 & \textbf{0.09} \\
\quad $w_b = 1.0$~s & 1.5 & 1.558 & 0.13 \\
\quad $\mathbf{w_b = 1.5}$~\textbf{s} & \textbf{2.0} & \textbf{1.662} & \textbf{0.18} \\
\quad $w_b = 2.5$~s & 3.0 & 1.681 & 0.27 \\
\quad $w_b = 3.5$~s & 4.0 & 1.689 & 0.36 \\
\midrule
\textit{(b) Hop $w_c$ ($w_b = 1.5$~s)} & & & \\
\quad $w_c = 0.10$~s & 1.6 & 1.674 & 0.71 \\
\quad $w_c = 0.25$~s & 1.75 & 1.669 & 0.31 \\
\quad $\mathbf{w_c = 0.50}$~\textbf{s} & \textbf{2.0} & \textbf{1.662} & \textbf{0.18} \\
\quad $w_c = 1.00$~s & 2.5 & 1.594 & \textbf{0.11} \\
\bottomrule
\end{tabularx}
\\ \vspace{6pt} 
{\footnotesize \noindent \raggedright The model was evaluated on the KUL dataset with the auditory attractor enabled. The total window is $w = w_b + w_c$. Bold text denotes the operating point used in Table~\ref{tab:online_comparision}. \par}
\end{table}

\vspace{-0.3cm}
\section{Conclusion}\label{sec:conclusion}

In this paper we introduce NeuroCodec, a generative EEG-guided target speaker extraction framework that predicts latent representations of a frozen high-fidelity audio codec instead of estimating time-frequency masks. By combining graph-based EEG modeling, cross-modal attention, and Mamba-based temporal modeling, NeuroCodec achieves effective EEG-audio fusion in a compact latent space. Experiments on two datasets demonstrate improved speech intelligibility and perceptual quality over existing approaches, as measured by both objective metrics and a subjective listening study. NeuroCodec also enables 44.1kHz reconstruction with spectral content extending to approximately 20~kHz with minimal speaker interference, unlike prior work with limited output bandwidth and speaker leakage. We extend the model to a causal streaming version, OnlineNeuroCodec, which enables streaming TSE using a causal auditory attractor derived from previous target estimates. This model achieves vastly superior computational speed while retaining competitive extraction quality. These results demonstrate the potential of generative latent modeling for neuro-steered speech extraction and future assistive hearing systems. However, OnlineNeuroCodec, like most other contemporary models, remains far from deployment on embedded hearing-aid hardware. Future work will explore autoregressive codec-token prediction for fully causal generation with reduced buffering requirements, potentially lowering algorithmic latency, alongside model compression and low-power implementation.

\section*{Acknowledgments}
This work was supported by the Accelerating Higher Education Expansion and Development (AHEAD) operation (Grant No. 6026-LK/8743-LK, World Bank). The authors also thank all participants who took part in the listening study.

\bibliographystyle{IEEEtran}
\bibliography{mybib}

\vfill

\end{document}